\documentstyle[epsf]{mn}
\input psfig

\begin{document}
\title[Ellipsoidal collapse and an improved model]
{Ellipsoidal collapse and an improved model for the number and spatial distribution of dark 
matter haloes}
\author[R. K. Sheth, H. J. Mo \& G. Tormen]
{Ravi K. Sheth$^1$, H. J. Mo$^1$ \& Giuseppe Tormen$^2$\\
$^1$ Max-Planck Institut f\"ur Astrophysik, 85740 Garching, Germany\\
$^2$ Dipartimento di Astronomia, 35122 Padova, Italy \\}
\date{Submitted 1 July 1999}

\maketitle

\begin{abstract}
The Press--Schechter, excursion set approach allows one to make 
predictions about the shape and evolution of the mass function of 
bound objects.  The approach combines the assumption that objects 
collapse spherically with the assumption that the initial density 
fluctuations were Gaussian and small.  While the predicted 
mass function is reasonably accurate at the high mass end, it has 
more low mass objects than are seen in simulations of hierarchical 
clustering.  We show that the discrepancy between theory and 
simulation can be reduced substantially if bound structures are 
assumed to form from an ellipsoidal, rather than a spherical 
collapse.  
In the original, standard, spherical model, a region collapses if 
the initial density within it exceeds a threshold value, 
$\delta_{\rm sc}$.  This value is independent of the initial size 
of the region, and since the mass of the collapsed object is related 
to its initial size, this means that $\delta_{\rm sc}$ is independent 
of final mass.  
In the ellipsoidal model, the collapse of a region depends on 
the surrounding shear field, as well as on its initial overdensity.  
In Gaussian random fields, the distribution of these quantities 
depends on the size of the region considered.  Since the mass of 
a region is related to its initial size, there is a relation between 
the density threshold value required for collapse, and the mass of 
the final object.  We provide a fitting function to this 
$\delta_{\rm ec}(m)$ relation which simplifies the inclusion of 
ellipsoidal dynamics in the excursion set approach.
We discuss the relation between the excursion set predictions 
and the halo distribution in high resolution N-body simulations, 
and use our new formulation of the approach to show that our simple 
parametrization of the ellipsoidal collapse model represents a 
significant improvement on the spherical model on an 
object-by-object basis.  Finally, we show that the associated 
statistical predictions, the mass function and the large scale 
halo-to-mass bias relation, are also more accurate than the 
standard predictions.  

\end{abstract}

\begin{keywords}  galaxies: clustering -- cosmology: theory -- dark matter.
\end{keywords}

\section{Introduction}\label{intro}
Current models of galaxy formation assume that structure grows 
hierarchically from small, initially Gaussian density fluctuations.  
Collapsed, virialized dark matter haloes condense out of the 
initial fluctuation field, and it is within these haloes that 
gas cools and stars form (White \& Rees 1977; White \& Frenk 1991; 
Kauffmann et al. 1999).  In such models, understanding the properties 
of these dark haloes is important.  There is some hope that dark 
haloes will be relatively simple to understand, because, to a good 
approximation, gravity alone determines their properties.  The 
formation and other properties of dark haloes can be studied using 
both N-body simulations and analytical models.  The most developed 
analytic model, at present, has come to be called the 
Press--Schechter approach.  It allows one to compute good 
approximations to the mass function (Press \& Schechter 1974; 
Bond et al. 1991), the merging history (Lacey \& Cole 1993, 1994; 
Sheth 1996; Sheth \& Lemson 1999b) and the spatial clustering 
(Mo \& White 1996; Mo, Jing \& White 1996, 1997; 
Catelan et al. 1998; Sheth 1998; Sheth \& Lemson 1999a) 
of dark haloes.  

Let $n(m,z)$ denote the number density of bound objects that 
have mass $m$ at time $z$.  Press \& Schechter (1974) argued that 
collapsed haloes at a late time could be identified with overdense 
regions in the initial density field.  Bond et al. (1991) described 
how the assumption that objects form by spherical collapse could be 
combined with fact that the initial fluctuation distribution was 
Gaussian, to predict $n(m,z)$.  
To do so, they made two assumptions:  
(i) a region collapses at time $z$ if the initial overdensity within 
it exceeds a critical value, $\delta_{\rm sc}(z)$.  This critical 
value depends on $z$, but is independent of the initial size of 
the region.  The dependence of $\delta_{\rm sc}$ on $z$ is 
given by the spherical collapse model.  
(ii) the Gaussian nature of the fluctuation field means 
that a good approximation to $n(m,z)$ is given by considering 
the barrier crossing statistics of many independent, uncorrelated 
random walks, where the barrier shape $B(m,z)$ is given by the 
fact that, in the spherical model, $\delta_{\rm sc}$ is independent 
of $m$.  

While the mass function predicted by this `standard' model is 
reasonably accurate, numerical simulations show that it may 
fail for small haloes (Lacey \& Cole 1994; Sheth \& Tormen 1999).  
This discrepancy is not surprising, because many assumptions must 
be made to arrive at reasonably simple analytic predictions.  
In particular, the spherical collapse approximation to the dynamics 
may not be accurate, because we know that perturbations in Gaussian 
density fields are inherently triaxial (Doroshkevich 1970; 
Bardeen et al. 1986).  

In this paper, we modify the standard formalism by incorporating 
the effects of non-spherical collapse.  
In Section~\ref{excursion} we argue that the main effect of 
including the dynamics of ellipsoidal rather than spherical 
collapse is to introduce a simple dependence of the critical 
density required for collapse on the halo mass.  There is some 
discussion in the literature as to why the excursion set approach 
works.  Section~\ref{model} continues this, and shows that our simple 
change to the `standard' model reduces the scatter between the 
predicted and actual masses of haloes on an object-by-object basis.  
Section~\ref{apply} shows that this simple change also 
substantially improves the agreement between predicted statistical 
quantities (the halo mass function and halo-to-mass bias relations) 
and the corresponding simulation results.  A final section summarizes 
our findings, and discusses how they are related to the work of 
Monaco (1995), Bond \& Myers (1996), Audit, Teyssier \& Alimi (1997), 
and Lee \& Shandarin (1998).  

\section{The excursion set approach}\label{excursion}
The first part of this section summarizes the `standard' 
model in which the spherical collapse model is combined 
with the assumption that the initial fluctuations were 
Gaussian and small.  The second part shows how the `standard' 
model can be modified to incorporate the effects of 
ellipsoidal, rather than spherical, collapse.  

\subsection{Spherical collapse:  the constant barrier}\label{constant}
Let $\sigma(r)$ denote the rms fluctuation on the scale $r$.  
In hierarchical models of clustering from Gaussian initial 
fluctuations, $\sigma$ decreases as $r$ increases in a way that 
is specified by the power spectrum.  If the initial fluctuations 
were small, then the mass $m$ within a region of size $r$ is 
just $m\propto r^3$.  
Bond et al. (1991) argued that the mass function of collapsed 
objects at redshift $z$, $n(m,z)$, satisfies 
\begin{equation}
\nu\,f(\nu)\equiv m^2\,{n(m,z)\over \bar\rho}\,
{{\rm d}\log m\over {\rm d}\log\nu},
\label{nufnu}
\end{equation}
where $\bar\rho$ is the background density, 
$\nu=\delta_{\rm sc}(z)/\sigma(m)$ is the ratio of the 
critical overdensity required for collapse in the spherical 
model to the rms density fluctuation on the scale $r$ of the initial 
size of the object $m$, and the function of the left hand side is 
given by computing the distribution of first crossings, 
$f(\nu)\,{\rm d}\nu$, of a barrier $B(\nu)$, by independent, 
uncorrelated Brownian motion random walks.  
Thus, in their model, for Gaussian initial fluctuations, 
$n(m,z)$ is determined by the shape of the barrier, $B(\nu)$, 
and by the relation between the variable $\nu$ and the mass $m$ 
(i.e., by the initial power spectrum).  

Bond et al. used the spherical collapse model to determine 
the barrier height $B$ as a function of $\nu$ as follows.  
In the spherical collapse model, the critical overdensity 
$\delta_{\rm sc}(z)$ required for collapse at $z$ is independent 
of the mass $m$ of the collapsed region, so it is independent of 
$\sigma(m)$.  Therefore, Bond et al. argued that since 
$\nu \equiv (\delta_{\rm sc}/\sigma)$, then $B(\nu)$ must be 
the same constant for all $\nu$.  
Using the spherical collapse model to set $\delta_{\rm sc}(z)$ 
means, e.g., that $B = \delta_{\rm sc}(z)=1.68647\,(1+z)$ in an 
Einstein-de~Sitter universe.  Since the barrier height 
associated with the spherical collapse model does not depend on 
$\nu=(\sigma_*/\sigma)$, and since the random walks are assumed to 
be independent and uncorrelated, the first crossing distribution 
can be derived analytically.  This allowed Bond et al. (1991) to 
provide a simple formula for the shape of the mass function that 
is associated with the dynamics of spherical collapse:  
\begin{equation}
\nu\,f(\nu) = 2\left({\nu^2\over 2\pi}\right)^{1/2}\ 
\exp\left(-{\nu^2\over 2}\right).
\label{oldps}
\end{equation}
Notice that in this approach, the effects of the background cosmology 
and power spectrum shape can be neatly separated.  The cosmological 
model determines how $\delta_{\rm sc}$ depends on $z$, whereas and 
the shape of the power spectrum fixes how the variance depends on 
scale $r$, so it fixes how $\sigma$ depends on mass $m\propto r^3$.  
Furthermore, for scale free spectra, if the mass function is well 
determined at one output time, then the others can be computed by 
simple rescalings.  

In this excursion set approach, the shape of the mass 
function is determined by $B(\sigma)$ and by $\sigma(m)$.  
Since $\sigma(m)$ depends on the shape of the initial power spectrum 
but not on the underlying dynamics, to incorporate the effects of 
ellipsoidal collapse into the excursion set model, we only need to 
determine the barrier shape associated with the new, non-spherical 
dynamics.  Below, we describe a simple way to do this.  

\subsection{Ellipsoidal collapse:  the moving barrier}\label{moving}
The gravitational collapse of homogeneous ellipsoids has been 
studied by Icke (1973), White \& Silk (1979), Peebles (1980), 
and Lemson (1993).  We will use the model in the form described 
by Bond \& Myers (1996).  That is, the evolution of the 
perturbation is assumed to be better described by the initial 
shear field than the initial density field, initial 
conditions and external tides are chosen to recover the Zeldovich 
approximation in the linear regime, and virialization is defined 
as the time when the third axis collapses.  
This last choice means that there is some freedom associated 
with how each axis is assumed to evolve after turnaround, and 
is the primary free parameter in the model we will describe below.  
Following Bond \& Myers (1996), we have chosen the following 
prescription.  Whereas, in principle, an axis may collapse to zero 
radius, collapse along each axis is frozen once it has shrunk by some 
critical factor.  This freeze-out radius is chosen so that 
the density contrast at virialization is the same (179 times the 
critical density) as in the spherical collapse model.  The results 
which follow are not very sensitive to the exact value of this 
freeze out radius.  

For a given cosmological background model (we will study the 
Einstein-de Sitter case in detail below), the evolution of an 
ellipsoidal perturbation is determined by three parameters:  
these are the three eigenvalues of the deformation tensor, 
or, equivalently, the initial ellipticity $e$, prolateness $p$, 
and density constrast $\delta$ (our $e$ and $p$ are what 
Bond \& Myers 1996 called $e_v$ and $p_v$, and are defined 
so that $|p|\le e$.  See Appendix~\ref{grfs} for details).  
Figure~\ref{dcplot} shows the expansion factor at collapse as a 
function of $e$ and $p$, for a region that had an initial 
overdensity $\delta=0.04215$, in an Einstein-de~Sitter universe.  
At a given $e$, the largest circles show the relation at $p=0$, 
medium sized circles show $|p|\le e/2$, and the smallest circles 
show $|p|\ge e/2$.  On average, virialization occurs later as $e$ 
increases, and, at a given $e$, it occurs later as $p$ decreases.  
For an Einstein-de~Sitter model the linear theory growth factor is 
proportional to the expansion factor, so this plot can be used to 
construct $\delta_{\rm ec}(e,p)$.  
For the range of $e$ and $p$ that are relevant for the results 
to follow, a reasonable approximation to this relation is given 
by solving 
\begin{equation}
{\delta_{\rm ec}(e,p)\over\delta_{\rm sc}} =  1 + \beta
  \left[5\,(e^2\pm p^2)\,
        {\delta^2_{\rm ec}(e,p)\over \delta^2_{\rm sc}}\right]^\gamma
\label{fitelcol}
\end{equation}
for $\delta_{\rm ec}(e,p)$, where $\beta=0.47$, 
$\gamma=0.615$, $\delta_{\rm sc}$ is the critical spherical 
collapse value, and the plus(minus) sign is used if $p$ is 
negative(positive).  [If $\gamma=0.5$ then this relation can 
be solved analytically to provide some feel for how $\delta_{\rm ec}$ 
depends on $e$ and $p$.  For example, when $\gamma=0.5$ and $p=0$, 
then $\delta_{\rm ec}\approx \delta_{\rm sc}/(1-e)$.]
The solid curve in Fig.~\ref{dcplot} shows the value given by 
equation~(\ref{fitelcol}) when $\gamma=0.615$ for $p=0$, and 
the two dashed curves show $|p|=e/2$.  

We want to consider the collapse of ellipsoids from an initially 
Gaussian fluctuation field.  Appendix~\ref{grfs} shows that on any 
scale $R_{\rm f}$ parameterized by $\sigma(R_{\rm f})$, there is 
a range of probable values of $e$, $p$ and $\delta$.  
This means that there is a range of collapse times associated with 
regions of size $R_{\rm f}$.  In principal, we could obtain an 
estimate for an average $\delta_{\rm ec}(\sigma)$ by averaging 
$\delta_{\rm ec}(e,p)$ over $p(e,p,\delta/\sigma)$ suitably.  
In essence, Monaco (1995), Audit, Teyssier \& Alimi (1997) and 
Lee \& Shandarin (1998) give different prescriptions for doing 
this.  We will use the simpler procedure described below.  

\begin{figure}
\centering
\mbox{\psfig{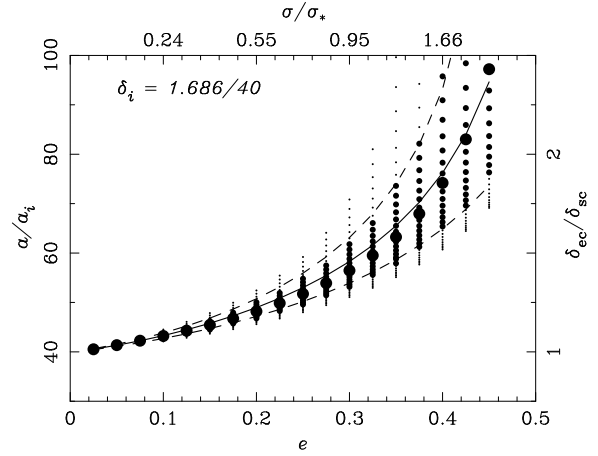}}
\caption{  The evolution of an ellipsoidal perturbation in an 
Einstein-de~Sitter universe.  Symbols show the expansion factor 
when the longest axis collapses and virializes, as a function of 
initial $e$ and $p$, in steps of 0.025, if the initial overdensity 
was $\delta_i$.  The solid curve shows our simple formula for the 
$p=0$ result, and the dashed curves show $|p|=e/2$.  The time 
required to collapse increases mononically as $p$ decreases.  The 
axis on the right shows the associated critical overdensity required 
for collapse, and the axis on the top shows the result of using our 
simple formula to translate from $e$ to $\sigma(m)$ when $p=0$.}
\label{dcplot}
\end{figure}

On average in a Gaussian field, $p=0$.  
The solid curve in Fig.~\ref{dcplot} shows the expansion factor 
at virialization in this case.  It is straightforward to use this 
curve to compute the associated $\delta_{\rm ec}(e,z)$.  
Having done so, if we can relate $e$ to the mass $m$, 
then we will be in a position to describe the barrier shape 
associated with ellipsoidal, rather than spherical collapse.  
This can be done as follows.  Regions initially having a given 
value of $\delta/\sigma$ most probably have an ellipticity 
$e_{\rm mp} = (\sigma/\delta)/\sqrt{5}$ (see Appendix).  To 
collapse and form a bound object at $z$, the initial overdensity 
of such a region must have been $\delta_{\rm ec}(e_{\rm mp},z)$.  
If we require that $\delta$ on the right hand side of 
this relation for $e_{\rm mp}$ be equal to this critical value 
$\delta_{\rm ec}(e_{\rm mp},z)$, then this sets 
$\sigma^2(R_{\rm f})$.  Since $R_{\rm f}^3$ is proportional 
to mass, this provides a relation between $e$ and mass, and so 
between $\delta_{\rm ec}$ and mass:   
\begin{equation}
\delta_{\rm ec}(\sigma,z) = \delta_{\rm sc}(z)\,
\left(1 + \beta \left[{\sigma^2\over \sigma^2_*(z)}\right]^\gamma\right),
\label{elcolbar}
\end{equation}
where we set $\sigma_*(z)\equiv\delta_{\rm sc}(z)$.  The axis labels 
on the top and right of the plot show this ($p=0$) relation.  

Notice that the power spectrum enters only in the relation between 
$\sigma$ and $m$, whereas the effects of cosmology enter only 
in the relation between $\delta_{\rm sc}$ and $z$.
For example, this expression is approximately the same for 
SCDM, OCDM, and $\Lambda$CDM models if all variances 
$\sigma^2(m)$ are computed using the model dependent power spectrum, 
and the value of $\delta_{\rm sc}(z)$ is computed using the 
spherical collapse model after including its dependence on 
background cosmology:  the differences between these 
models arise primarily from converting the scaling variable $\nu$ 
to the physical variables $z$ and $m$.  

A number of features of equation~(\ref{elcolbar}) are worth 
noticing.  Massive objects have $\sigma/\sigma_*\ll 1$.  For such 
objects equation~(\ref{elcolbar}) suggests that 
$\delta_{\rm ec}(\sigma,z)\approx \delta_{\rm sc}(z)$, 
so the critical overdensity required for collapse at $z$ is 
approximately independent of mass:  massive objects are well 
described by the spherical collapse model.  
Other approaches yield the same result (e.g., Bernardeau 1994).  
Second, the critical overdensity increases with $\sigma(m)$, 
so it is larger for less massive objects.  
This is because smaller objects are more influenced by external 
tides; they must have a greater internal density if they are to 
hold themselves together as they collapse.  

Equation~(\ref{elcolbar}) is extremely useful because it 
allows one to include the effects of ellipsoidal collapse 
into the Bond et al. (1991) excursion set model in a 
straightforward manner.  Namely, all we need to do is to 
use equation~(\ref{elcolbar}) when setting 
$B(\sigma,z)=\delta_{\rm ec}(\sigma,z)$.  
Then the distribution of first crossings of this barrier by 
independent random walks can be used to give an estimate of 
the mass function associated with ellipsoidal collapse.  
For example, it is straightforward to simulate an ensemble of 
independent unconstrained random walks, and to record the 
distribution of first crossings of the ellipsoidal collapse 
`moving' barrier.  To a very good approximation, this first 
crossing distribution is 
\begin{equation}
\nu\,f(\nu) = 2A\,\left(1 + {1\over \nu{^{2q}}}\right)\ 
\left({\nu{^2}\over 2\pi}\right)^{1/2}\,\exp\left(-{\nu{^2}\over 2}\right) ,
\label{ftell}
\end{equation}
where $\nu$ was defined earlier, $q=0.3$ and $A\approx 0.3222$.  
This first crossing distribution differs from that predicted by 
the `standard' constant barrier model (equation~\ref{oldps}) for 
which $q=0$ and $A=1/2$.  

The great virtue of interpreting equation~(\ref{elcolbar}) 
as the `moving' barrier shape is that, once the barrier shape is 
known, all the predictions of the excursion set program can be 
computed relatively easily.  
This means that we can use the logic of Lacey \& Cole (1993) to 
compute the conditional mass functions associated with ellipsoidal 
rather than spherical collapse.  As in the original 
model, this is given by considering the successive crossing of 
boundaries associated with different redshifts.  Once this 
conditional mass function is known, the forest of merger history 
trees can be constructed using the algorithm described by 
Sheth \& Lemson (1999b), from which the nonlinear stochastic 
biasing associated with this mass function can be derived using 
the logic of Mo \& White (1996) and Sheth \& Lemson (1999a).  

\section{Excursion set predictions and N-body simulations}\label{model}
The mass function in equation~(\ref{oldps}) was first derived 
by Press \& Schechter (1974).  They used the Gaussian statistics 
of regions which are denser than $\delta_{\rm sc}(z)$ on a given 
scale $\sigma(m)$ to compute the mass function of haloes at redshift 
$z$.  However, their derivation did not properly address what happens 
to regions which are denser than $\delta_{\rm sc}(z)$ on more than one 
scale.  The excursion set approach of Bond et al. (1991) shows how 
to do this.  It is based on the following hypothesis:  at $z$, 
the mass of a collapsed object is the same as the mass within the 
largest region in the initial conditions that could have collapsed 
at $z$.  

Unfortunately, this hypothesis makes no reference to the centre 
of the collapsed object, either in the initial conditions or at 
the final time, whereas the Bond et al. calculation does.  
This has led to some discussion in the literature as to exactly 
how one should compare the excursion set approach predictions with 
the  haloes which form in numerical simulations of hierarchical 
clustering.  These discussions have led to the perception that, on 
an object-by-object basis, the excursion set predictions are 
extremely unreliable (Bond et al. 1991; White 1996), so that it is 
difficult to explain why, in a statistical sense, the excursion set 
predictions work as well as they do (Monaco 1999).  
This section provides a discussion of how the predictions of this 
approach are related to the results of numerical simulations.  
It then shows that the excursion set approach does, in fact, make 
accurate predicitions, even on an object-by-object basis.  This 
comparison shows that, on an object-by-object basis, our 
parametrization of ellipsoidal dynamics represents a 
significant improvement on the standard spherical model.  

\subsection{Selecting haloes in the initial conditions}\label{select}
Suppose that our statement of the excursion set hypothesis is correct:  
the largest region in the initial conditions that can collapse, will.  
Then it should be possible to combine the spherical collapse 
model with the statistics of the initial fluctuation field to 
obtain an estimate of the mass function of haloes at $z$.  
The natural way to do this is as follows.  
Generate the initial Gaussian random fluctuation field.  
Compute the average density within concentric spherical regions 
centred on each position of the field.  These are the excursion set 
trajectories associated with each position.  
At each position, find the largest spherical region within which the 
initial average density fluctuation exceeds $\delta_{\rm sc}(z)$.  
Call the mass within this region the predicted mass.  
Thus, for each position in the initial field, there is an associated 
$m_{\rm pred}(z)$.  
Go to the position with the largest $m_{\rm pred}(z)$, 
call this position $r_1$ and set $m_1=m_{\rm pred}$.  
Associated with $m_1$ is a spherical volume $v_1 = m_1/\bar\rho$ 
centred on $r_1$.  
Disregard the predicted masses (i.e. ignore the excursion set 
trajectories) for all the other positions within this $v_1$.  
If the simulation box has volume $V$, consider the remaining volume 
$V-v_1$.  Set $m_2$ equal to the largest value of $m_{\rm pred}(z)$ 
in the remaining volume $V-v_1$, and record this position $r_2$.  
Disregard the predicted mass for all other positions within the 
associated $v_2$.  Continue until the remaining volume in the 
simulation box is as small as desired.  
The resulting list of $m_i$s represents the halo mass function 
predicted by the excursion set approach.  
The list of positions $r_i$ represents the Lagrangian space 
positions of the haloes.  This is essentially the algorithm 
described at the end of Section~3.3 in Bond \& Myers (1996).  
(They also describe what to do in the event that, for example, 
some of the mass associated with $v_2$ was within $v_1$.)
Inclusion of ellipsoidal, rather than spherical dynamics, into 
this excursion set algorithm is trivial:  simply replace 
$\delta_{\rm sc}$ with $\delta_{\rm ec}(m)$.  

Although this algorithm follows naturally from the excursion set 
hypothesis, in practice, it is rather inefficient.  
For this reason, making a preliminary selection of candidate 
positions for the excursion set $r_i$s may be more efficient.
For example, whereas the algorithm described above selects peaks 
in the initial $m_{\rm pred}$ distribution, the positions of 
these peaks may correspond to peaks in the density field itself.  
Since these may be easier to identify, it may be more efficient 
to use them instead.  
Essentially, this is the motivation behind the peak--patch 
approach of Bond \& Myers (1996).  

\subsection{Predicted and actual halo masses}\label{pred}
\begin{figure*}
\centering
\mbox{\psfig{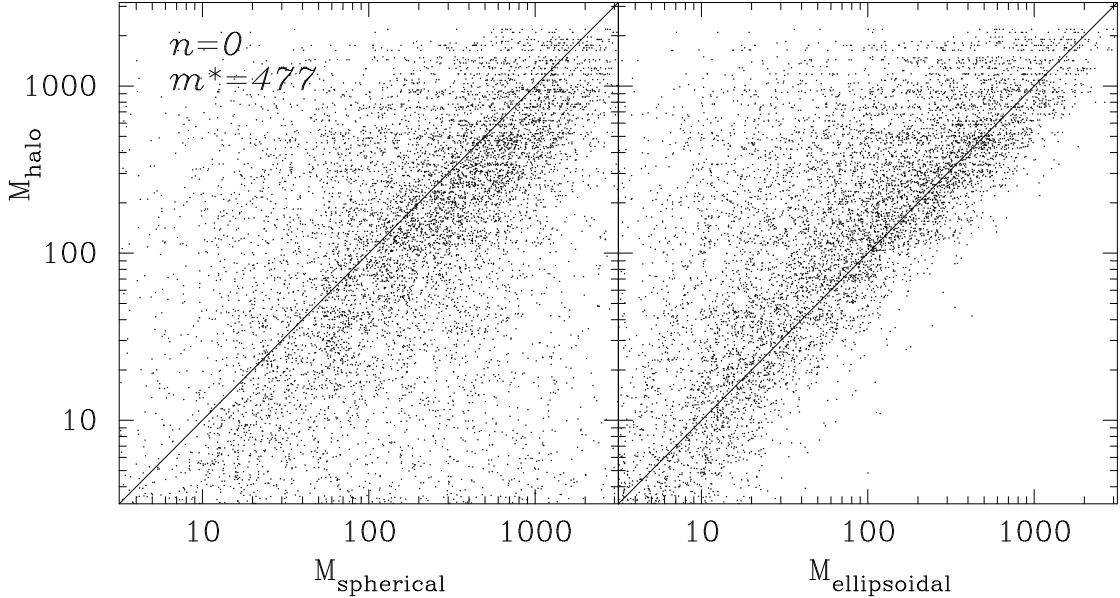}}
\caption{The mass of the halo in which a randomly chosen particle 
is, $M_{\rm halo}$, is plotted versus the mass predicted by the 
spherical (left panel) and ellipsoidal collapse (right panel) 
models.  A randomly chosen $10^4$ of the $10^6$ particles in a 
simulation of an Einstein-de Sitter universe with white noise 
initial conditions were used to make the plot.}
\label{random}
\end{figure*}

The algorithm described above shows that the only values of 
$m_{\rm pred}$ that are relevant are those that are in the list 
of $m_i$s.  That is, only a few stalks in the bundle of excursion 
set trajectories are actually associated with collapsed 
objects.  It is easy to understand why.  
Imagine running a numerical simulation.  Choose a random particle 
in the simulation, and record the mass $m$ of the halo in which 
this particle is at some specified redshift $z$.  Since the 
particle was chosen at random, it is almost certainly not the 
centre-of-mass particle of the halo in which it is at $z$.  
Is there a simple reason why the halo collapsed around the 
centre-of-mass particle, and not around the one chosen at random?  
The excursion set answer to this question is ``yes'':  
collapse occurs around positions which are initially local maxima 
of the excursion set predicted mass.  When collapse occurs, the 
approach assumes that shells do not cross, so initially concentric 
shells remain concentric.  
This means that the centre-of-mass particle at the final time 
is also the centre-of-mass particle initially (particles retain 
the binding energy ranking they had in the initial conditions), 
and that the predicted mass for this centre-of-mass particle is 
higher than for the one chosen at random.  This has the important 
consequence that only the centre-of-mass particle prediction is a 
good estimate for the mass of the halo at $z$; all other particles 
provide underestimates of the final mass.  

\begin{figure*}
\centering
\mbox{\psfig{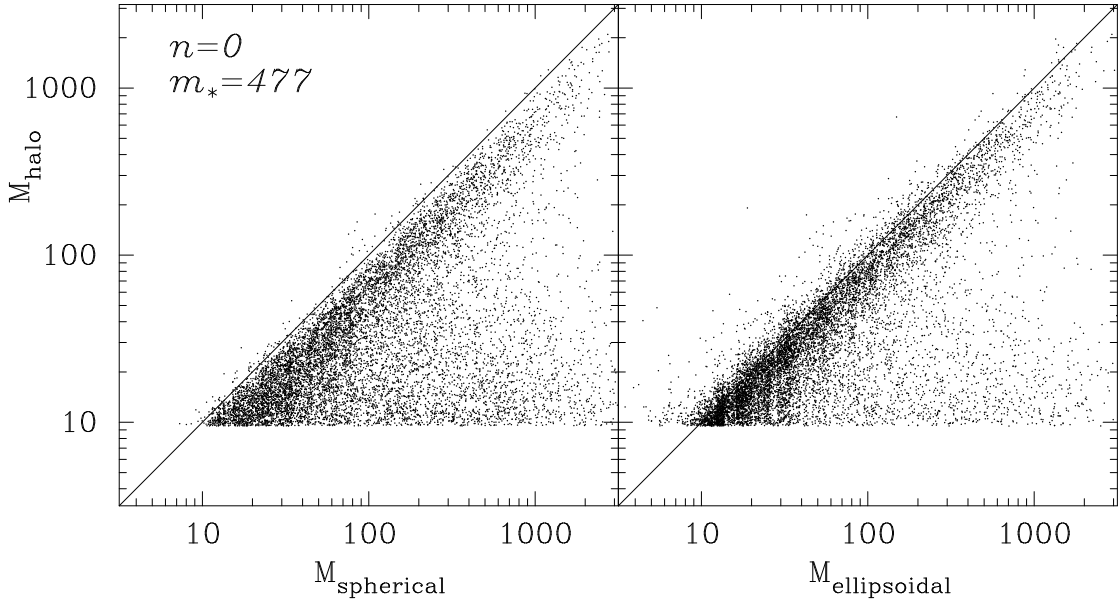}}
\caption{  The mass of a halo in a simulation of an 
Einstein-de Sitter universe with white noise initial conditions 
versus that predicted by the excursion set approach.  
The panel on the left shows the prediction associated with the 
`standard' spherical  collapse approximation to the dynamics; 
the panel on the right shows the prediction associated with our 
moving barrier parametrization of the ellipsoidal collapse model.  }
\label{centres}
\end{figure*}

\begin{figure}
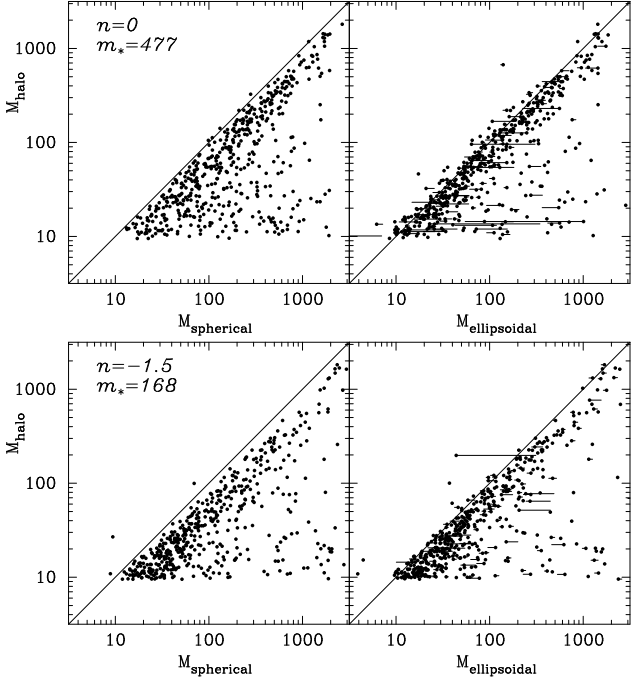

\centering
\mbox{\psfig{figure=ecmodelf4a.ps,height=4.5cm,bbllx=6pt,bblly=71pt,bburx=683pt,bbury=435pt}}
\mbox{\psfig{figure=ecmodelf4b.ps,height=4.5cm,bbllx=6pt,bblly=71pt,bburx=683pt,bbury=435pt}}
\caption{The effect of changing $p$ at a given $e$ on the predicted 
mass of a halo:  as $p$ becomes more negative(positive), 
 $\delta_{\rm ec}(e,p)$ increases(decreases), 
 so the predicted mass decreases(increases).  
The filled circles show the $p=0$ prediction used to produce the 
previous figure, and the bars show the range $|p|=0.33\,e$.  
The two panels on the top show the result for white noise initial 
conditions, and the bottom panels were constructed from simulations 
in which the slope of the initial power spectrum was $n=-1.5$.}
\label{errors}
\end{figure}

We will use Figures~\ref{random}--\ref{errors} to demonstrate this 
in two steps.  First, we will use Figure~\ref{random} to argue 
that ellipsoidal dynamics represents a significant improvement over 
the spherical model.  Then, we will use Figures~\ref{centres} 
and~\ref{errors} to show that our moving barrier excursion set 
approach associated with ellipsoidal dynamics allows one to make 
accurate predictions on an object-by-object basis.  These figures 
were constructed using numerical simulations which were kindly 
made available by Simon White, and are described in White (1996).  
They follow the clustering of $10^6$ particles from white noise 
initial conditions (of course, the results to follow are similar 
for other initial power spectra).  
We have chosen to show results for that output time ($a/a_i=36$) 
in the simulations in which the number of haloes containing more 
than ten particles each was $\sim 10^4$.  This number was chosen 
for ease of comparison with Fig.~8 of White (1996).

To show that the evolution of an object is well described by 
spherical or ellipsoidal dynamics, we should compare the 
evolution of the object's three axes with that of the model.  
For the spherical model, this has been done  by Lemson (1995).  
We will perform a cruder test here.  In the spherical collapse 
model, an object forms at $z$ if the initial overdensity 
within it exceeds $\delta_{\rm sc}(z)$.  Since, in the model, 
shells do not cross, so initially concentric regions remain 
concentric, we can compare $M_{\rm predicted}$, the mass contained 
within the largest spherical region centred on a randomly chosen 
particle in the initial conditions within which the density 
exceeds $\delta_{\rm sc}(z)$, with $M_{\rm halo}$, the mass of 
the object in which that particle actually is at $z$.  
The comparison with ellipsoidal dynamics is similar, except 
that one uses $\delta_{\rm ec}(M_{\rm halo})$, instead of the 
spherical collapse value, to compute the predicted mass.  
Thus, rather than testing the detailed evolution of the object, 
this simply tests whether or not the time it takes before 
virialization occurs depends on the initial overdensity in the 
way the model describes.  

Figure~\ref{random} shows this comparison for $10^4$ particles 
chosen randomly from the simulation.  (We use the same set of 
particles in both panels.  For cosmetic reasons, the predicted 
mass has been shifted randomly within each mass bin as described 
by White.)  
The panel on the left shows the scatter plot associated with 
spherical dynamics (it should be compared with White's plot, 
which was constructed from a simulation with $n=-1$ initial 
conditions), and the panel on the right shows the result of 
using our parametrization of ellipsoidal dynamics instead.  
Namely, the $y$ position associated with a particle is given by 
$M_{\rm halo}$, the mass of the halo in which the particle is, 
and the $x$ position is obtained as we described above.  

The difference between the two panels is striking:  
the points in the panel on the right populate the upper left 
half only.  This difference is easily understood:  
whereas, $\delta_{\rm sc}$ is independent of $M_{\rm halo}$, 
$\delta_{\rm ec}(m)$ increases as $m$ decreases.  Therefore, 
relative to the spherical model, the largest filter size 
containing the critical ellipsoidal collapse overdensity 
decreases as $M_{\rm halo}$ decreases, so that 
$M_{\rm ellipsoidal}\le M_{\rm spherical}$ always.  
Thus, in effect, including ellipsoidal dynamics moves all the 
points in the spherical model scatter plot to the left, and, 
on average, this shift depends on $M_{\rm halo}$.  

White (1996) argued that such a scatter plot could be used 
to test the excursion set approach.  He argued that if the 
Bond et al. (1991) formulation of the excursion set approach is 
correct, then there should be no scatter in such a plot.  
Figure~\ref{random} shows that, although the correlation between 
$M_{\rm halo}$ and $M_{\rm predicted}$ is tighter in the ellipsoidal 
than in the spherical model, the scatter in both panels is still 
considerable.  That this scatter is, in fact, quite 
large led White to argue that the accuracy of the excursion set 
predictions was surprising.  

However, as we discussed above, much of this scatter is a 
consequence of choosing random particles to construct the scatter 
plot.  We argued that because random particles will almost always 
provide an underestimate of the true mass, such a plot should be 
populated only in the upper left half.  This is clearly not the 
case for spherical dynamics (the panel on the left).  
Whereas the panel on the right looks more like we expect, it is 
not really a fair test of the ellipsoidal collapse, moving barrier, 
excursion set model, because it was constructed using a fixed 
$\delta_{\rm ec}(M_{\rm halo})$, rather than one which depends 
on scale, to compute $M_{\rm predicted}$.  Using the scale 
dependent $\delta_{\rm ec}(m)$ relation, rather than the fixed 
value $\delta_{\rm ec}(M_{\rm halo})$, to construct the plot 
will have the effect of moving some of the points to the right.  
Nevertheless, this panel suggests that inclusion of ellipsoidal 
dynamics represents a significant improvement over the spherical 
model.  

To make this point more clearly, Fig.~\ref{centres} shows the 
scatter plot one obtains by using only those particles which are 
centres of haloes to make the comparison between theory and 
simulations.  (Only haloes containing more than 10 particles 
were used to make this plot, since discreteness effects in the 
initial conditions become important on the small scales initially 
occupied by less massive haloes).  As before, the panel on the 
left shows the result of using spherical dynamics to compute the 
predicted mass, and the one on the right shows the one associated 
with ellipsoidal dynamics---but now, the predicted mass is 
computed using the ellipsoidal collapse moving barrier, rather 
than one fixed at the value associated with $M_{\rm halo}$.  
The most striking difference between this plot and the previous 
one is that now the upper left half in both panels is empty.  
As we discussed above, this provides strong support for our 
excursion set assumption that collapse occurs around local maxima 
of the $m_{\rm pred}$ distribution.  

In addition to showing the correlation, on an object-by-object basis, 
between predicted and simulated masses  more clearly, 
using only the centre of mass particles when constructing the 
scatter plot allows us to test the relative merits of the 
spherical and ellipsoidal model approximations to the exact 
dynamics.  In both panels, some of the discrepancy between 
prediction and simulation arise if some of the mass predicted 
to be in a halo was already assigned to a halo of larger mass, 
because $M_{\rm pred}>M_{\rm halo}$ produces points which populate 
the bottom right half of the plot.  
However, in the ellipsoidal model, some of the discrepancy 
almost certainly arises from the fact that we use a very simple 
prescription for relating the mass to $e$ and $p$.  
Presumably, the scatter in the panel on the right can be reduced 
by explicitly computing $\delta_{\rm ec}(e,p)$, and using this 
to compute $M_{\rm ellipsoidal}$, rather than by using the 
representative value $e_{\rm mp}$ that we adopted when deriving 
equation~(\ref{elcolbar}).  

Fig.~\ref{errors} shows the result of accounting for the effects 
of this scatter in the following crude way.  
The initial region containing the mass of $M_{\rm halo}$ could 
have had different values of $e$ and $p$ than the ones we assumed.  
Since $\delta_{\rm ec}$ is a function of $e$ and $p$, changing 
these values results in a different predicted $M_{\rm ellipsoidal}$.  
The lines through each point in the figure illustrate the range of 
predicted masses associated with each object if the ellipsoidal 
collapse barrier in equation~(\ref{elcolbar}) had 
$|p|= \pm 0.33\,e$.  On any given scale, integrating 
$g(e_{\rm mp},p|\delta)$ over this range in $p$ 
(recall $e_{\rm mp} = (\sigma/\delta)/\sqrt{5}$), 
shows that $p$ falls in this range approximately $50\%$ of the 
time (and in the range $|p|/e=0.5$, $70\%$ of the time).  
For clarity, of the $\sim 10^4$ objects, only a randomly chosen 
five hundred are shown.  The plot shows the correlation more 
clearly than the previous figure.  It also shows that, at least for 
some of the objects, the difference between the predicted and 
actual masses may be attributed to the scatter in initial values 
of $e$ and $p$.  We have not pursued this in further detail.  

We feel that, taken together, the three figures above make two 
points.  Firstly, because the upper left half of the scatter plot 
for centre-of-mass particles really is empty, our excursion set 
hypothesis that collapse occurs around local maxima of the 
$m_{\rm pred}$ distribution must be quite accurate.  
Secondly, because the centre-of-mass points follow the 
$M_{\rm halo}=M_{\rm predicted}$ relation reasonably well, 
and because the scatter around this mean relation is smaller for 
the ellipsoidal than for spherical dynamics predictions, our 
parametrization of ellipsoidal dynamics in the excursion set 
approach represents a significant improvement on the spherical 
model, on an object-by-object basis.  

We also think it important to point out that our model requires 
that collapse have occured along all three axes.  Had we chosen 
collapse along only the first axis to signify virialization, 
$\delta_{\rm ec}(m)$ would decrease with $m$.  In this case, 
$M_{\rm ellipsoidal}\ge M_{\rm spherical}$, and including ellipsoidal 
dynamics would increase the scatter in Fig.~\ref{random}.  
Moreover, all points in the left panel of Fig.~\ref{centres} 
would be shifted to the right, with points having small 
$M_{\rm halo}$ being shifted further.  Thus, if there were any 
correlation between the predicted and simulated masses in the 
resulting scatter plot, it would not be along the 
$M_{\rm halo}=M_{\rm predicted}$ line.  Therefore, 
Figs.~\ref{random}--\ref{errors} provide strong empirical 
justification for our identification of virialization with the 
time at which all three axes of the initial ellipsoid collapse.  

Finally, because the centre-of-mass particles really do show the 
expected correlation, if one is interested in studying the 
statistical properties of collapsed objects, then it should be a 
good approximation to study only these centre-of-mass particles.  
For example, suppose one is interested in using the fact that 
the initial distribution was a Gaussian random field to predict 
the fraction of mass which is contained in objects which 
have collapsed along all three axes.  Since only 8\% of all 
positions in an initial Gaussian field are predicted to collapse 
along all three axes (e.g. Lee \& Shandarin 1998) one might 
conclude that only 8\% of the mass can be contained in such 
objects.  However, just as the only relevant excursion set 
predictions are those associated with centre-of-mass trajectories 
(in our model the excursion set trajectories and associated 
predictions centred on other positions, while almost surely 
wrong, are irrelevant), so also is this value of 8\%, because it is 
based on the statistics of random positions, a very misleading number.  
The relevant question is not what fraction of all positions can 
collapse along all three axes, but what fraction of centre-of-mass 
particles (or, equivalently, peaks in the initial $m_{\rm pred}$ 
distribution) can collapse along all three axes.  This fraction 
is almost certainly closer to unity than to 8\%.  Moreover, since 
each such particle may be at the centre of a collapsed halo that 
has a mass considerably greater than that of a single particle, 
the actual fraction of mass that is in objects that have 
collapsed along all three axes can be considerable.  Since these 
centre-of-mass particles are almost certainly not randomly placed 
in the initial field, this fraction is more difficult to estimate, 
though it is certainly considerably greater than 8\%.  

This is also why computing other statistical quantities, such as 
the mass function of collapsed objects, is more complicated.  
Substituting the first crossing distribution associated with 
the ellipsoidal collapse moving barrier (equation~\ref{ftell}) 
in equation~(\ref{nufnu}) to compute the mass function is 
equivalent to assuming that the statistics of randomly chosen 
particles are the same as those of centre-of-mass particles.  
This is not an unreasonable first approximation.  
(We plan to present a more detailed derivation of the relation 
between the first crossing distribution associated with independent 
random excursion set trajectories and the mass function associated 
with centre-of-mass trajectories in a separate paper.  The more 
detailed derivation shows that this simple approximation is also 
reasonably accurate.)  
This approximation gives the original Press-Schechter, 
Bond et al. (1991) formula for the mass function associated with 
spherical collapse, and equation~(\ref{ftell}) for the mass function 
associated with our parametrization of ellipsoidal collapse.  

\section{Statistical predictions}\label{apply}
This section provides two examples of the increase in accuracy 
of the predicted statistical quantities that results from the 
inclusion of ellipsoidal dynamics in the excursion set approach.  

\subsection{The mass function}\label{newnm}
Fig.~2 of Sheth \& Tormen (1999) shows that, in the GIF 
(Kauffmann et al. 1999) simulations of clustering in 
SCDM, OCDM and $\Lambda$CDM models, the unconditional mass 
function is well approximated by 
\begin{equation}
\nu\,f(\nu) = 2A\,\left(1 + {1\over \nu'{^{2q}}}\right)\ 
\left({\nu'{^2}\over 2\pi}\right)^{1/2}\,\exp\left(-{\nu'{^2}\over 2}\right) ,
\label{giffit}
\end{equation}
where $\nu'=\sqrt{a}\,\nu$, $a=0.707$, $q=0.3$ and $A\approx 0.322$ 
is determined by requiring that the integral of $f(\nu)$ over all 
$\nu$ give unity (this last just says that all the mass is 
assumed to be in bound objects of some mass, however small).  
Essentially, the factor of $a=0.707$ is determined by the 
number of massive haloes in the simulations, and the parameter $q$ 
is determined by the shape of the mass function at the low mass 
end.  The GIF mass function differs from that predicted by the 
`standard' model (equation~\ref{oldps}) for which $a=1$, 
$q=0$, and $A=1/2$.  The simulations have more massive haloes 
and fewer intermediate and small mass haloes than predicted by 
equation~(\ref{oldps}).  Comparison with equation~(\ref{ftell})
shows that the two expressions are identical, except for the 
factor of $a$.  

To show this more clearly, we can derive numerically (following 
Sheth 1998) the shape of the barrier $B(\sigma,z)$ which gives rise 
to the GIF mass function of equation~(\ref{giffit}), if the 
relation between the first crossing distribution 
$f(\sigma)\,{\rm d}\sigma$ of independent unconstrained Brownian 
walks and the halo mass function is given by equation~(\ref{nufnu}).  
Since the random walk problem can also be phrased in terms of 
the scaled variable $\nu$, and since the GIF mass functions can 
also be expressed in this variable, we only need to compute the 
barrier shape once; simple rescaling of the variables gives the 
barrier shape at all later times.  
To a very good approximation, the barrier associated with the GIF 
simulations has the form 
\begin{equation}
B_{\rm GIF}(\sigma,z) = \sqrt{a}\,\delta_{\rm sc}(z)\,
\left(1 + b \left[{\sigma^2\over a\,\sigma^2_*(z)}\right]^c\right) ,
\label{gifbar}
\end{equation}
where $\delta_{\rm sc}(z)=\sigma_*(z)$, $\sigma(m)$ and $a$ are 
the same parameters that appear in the mass function, so 
$\delta_{\rm sc}(z)$ is given by the spherical collapse model and 
depends on the cosmological model, $\sigma(m)$ depends on the shape 
of the initial fluctuation spectrum, 
$\sigma/\sigma_*(z) \equiv \sigma(m)/\delta_{\rm sc}(z)\equiv 1/\nu$, 
$b=0.5$, and $c = 0.6$.  
Notice that this barrier shape (equation~\ref{gifbar}) which is 
required to yield the GIF mass function (equation~\ref{giffit}) 
has the same functional form as the barrier shape associated with 
the ellipsoidal collapse model (equation~\ref{elcolbar}).  
Except for the factor of $a$, the two barriers are virtually 
identical.  

To some extent, the value of $a$ is determined by how the haloes 
were identified in the simulations.  
There is some freedom in how one this is done.  
Typically, one uses a friends-of-friends or spherical 
overdensity algorithm to identify bound groups.  Both algorithms 
have a free parameter which is usually set by using the spherical 
collapse model.  In the spherical overdensity case, the overdensity 
is usually set to $\sim 200$ times the background density.  In the 
friends-of-friends case, it is customary to set the link-length 
to 0.2 times the mean interparticle separation.  Clearly, the 
shape of the mass function will depend on how groups are 
identified.  In the friends-of-friends case, for example, 
decreasing the link-length will result in fewer massive objects.  
Since we are considering the mass function associated with 
collapsed ellipsoids, it is not obvious any more that the free 
parameters in these group finders should be set using the spherical 
collapse values.  

Consider what happens as we change the link-length in the 
friends-of-friends case.  If, on average, the density profile of the 
objects identified using a given link-length is a power law, then 
decreasing the link-length means that all haloes will become less 
massive by some multiplicative factor.  If this power law is 
approximately independent of halo mass, then this factor will also 
be approximately independent of halo mass.  This means that, for 
some range of scales, there is a degeneracy between the 
friends-of-friends link-length and the parameter $M_*$.  Since the 
mass function in the simulations is a function of $\sigma/\sigma_*$, 
this will translate into a degeneracy between the link length and 
$M_*$, so the degeneracy between link length and $\sigma_*$ may 
depend on power spectrum.  
For this reason, we will treat the parameter $a$ in 
equation~(\ref{giffit}) above as being related to the 
link-length.  The value $a = 0.707$ is that associated 
with a link-length which is 0.2 times the mean interparticle 
separation, the value suggested by the spherical collapse model,
when the power spectrum is from the CDM family.  
Presumably, if we were to decrease this link-length sufficiently, 
we would find $a\approx 1$.  Since the link-length associated 
with $a = 0.707$ is more or less standard, we have not 
changed it and recomputed the simulation mass function.  Rather, 
we have simply chosen to argue that the fact that the GIF barrier 
(equation~\ref{gifbar}) is simply a scaled version of the moving 
barrier of equation~(\ref{elcolbar}) argues strongly 
in support of the accuracy of the ellipsoidal collapse model.

\subsection{Biasing on large scales}\label{bias}
\begin{figure}
\centering
\mbox{\psfig{figure=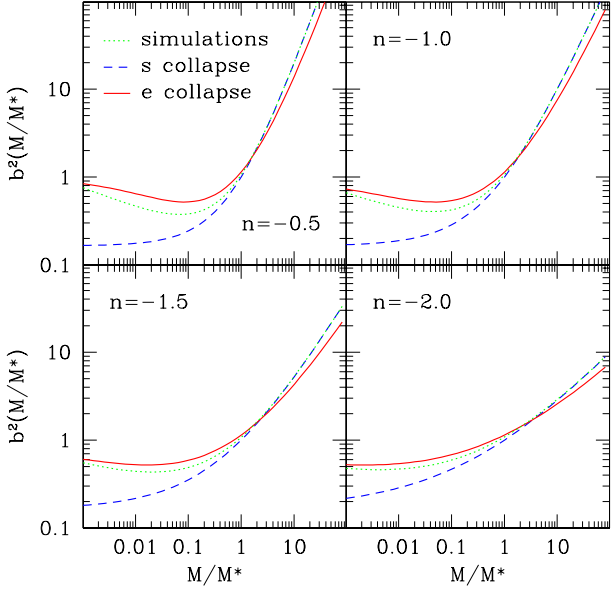,height=8.5cm}}
\caption{The large scale bias factor $b(m)$ as a function of halo 
mass.  Dotted curves show a fit to this relation measured 
in numerical simulations by Jing (1998), though his Figure~3 
shows that the bias factor for massive haloes in his simulations 
is slightly smaller than the one given by his fitting function.  
Dashed curves show the spherical collapse prediction of 
Mo \& White (1996), and solid curves show the elliposidal collapse 
prediction of this paper.  At the high mass end, our solid curves 
and the simulation results differ from Jing's fitting function 
(dotted) in the same qualitative sense. }
\label{scfbias}
\end{figure}

The large scale halo-to-mass bias relation in simulations is 
also different from that predicted by the `standard' spherical 
collapse model (Jing 1998, 1999; Sheth \& Lemson 1999a; 
Porciani, Catelan \& Lacey 1999; Sheth \& Tormen 1999).  
Fig.~\ref{scfbias} shows this large scale bias relation as a 
function of halo mass for haloes which form from 
initially scale free Gaussian random density fluctuation fields:  
i.e., the initial power spectrum was $P(k)\propto k^n$.  
The dotted line shows Jing's (1998) fit to this bias relation, 
measured in numerical simulations of hierarchical clustering.  
The dashed line shows the bias relation computed by 
Mo \& White (1996) using the `standard' spherical 
collapse, constant barrier model.  While it is reasonably accurate 
at the high mass end, the less massive haloes in the simulations 
appear to cluster more strongly than this model predicts.  
Sheth \& Tormen (1999) argued that some of this disrepancy arises 
from the fact that the mass function in the simulations differs 
from the Press--Schechter function.  They combined the simulation 
mass function with the peak background split approximation to 
estimate the large scale bias.  
If the rescaled mass function in Jing's scale free simulations 
is the same as that in the GIF simulations, then their peak 
background split formula fares better than the standard model, 
though it does not produce the upturn at low masses that Jing finds.
Moreover, Sheth \& Tormen gave no dynamical justification for 
why the mass function differs from the standard one.  

\begin{figure}
\centering
\mbox{\psfig{figure=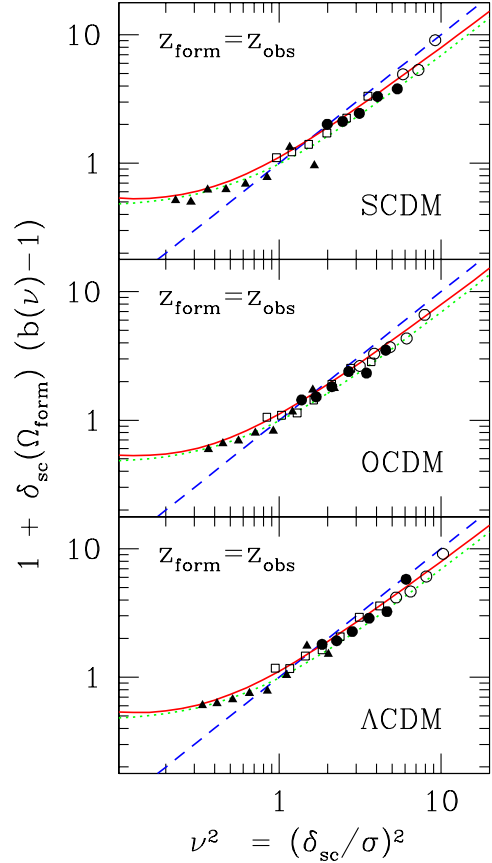,height=12cm,bbllx=18pt,bblly=145pt,bburx=335pt,bbury=700pt}}
\caption{The large scale bias factor $b(m)$ as a function of halo 
mass in the GIF simulations.  Dashed curves show the spherical 
collapse prediction of Mo \& White (1996), 
dotted curves show the peak background split formula 
of Sheth \& Tormen (1999), and solid curves show the ellipsoidal 
collapse prediction of this paper.  }
\label{gifbias}
\end{figure}

To compute the large scale bias relation associated 
with our ellipsoidal collapse, moving barrier model we must relate 
the bias relation to the random walk model.  This was been done by 
Mo \& White (1996), who argued that the bias relation was related 
to the crossing of two barriers (also see Sheth \& Tormen 1999).  
Essentially, the large scale bias relation is associated with 
random walks which travel far from the origin before intersecting 
the barrier.  To insure that this happens, one must consider random 
walks which intersect the barrier when the barrier height is very 
high.  We have simulated random walks, and recorded the first crossings 
of the barrier given in equation~(\ref{gifbar}) in the high-barrier 
limit.  We have then used the relation given by Mo \& White to 
compute the associated prediction for the large scale bias relation.  
To a very good approximation, this relation is 
\begin{displaymath}
b_{\rm Eul}(\nu) = 1 + b_{\rm Lag}(\nu),
\end{displaymath}
where $\nu \equiv \delta_{\rm sc}(z)/\sigma(m,z)$, and 
\begin{eqnarray}
b_{\rm Lag}(\nu) &=& {1\over\sqrt{a}\delta_{\rm sc}(z)} \ 
\Bigl[ \sqrt{a}\,(a\nu^2) + \sqrt{a}\,b\,(a\nu^2)^{1-c} \nonumber \\
&&\qquad\qquad\ \ - {(a\nu^2)^c\over (a\nu^2)^c + b\,(1-c)(1-c/2)}\Bigr] ,
\label{lagbias}
\end{eqnarray}
where $a$, $b$ and $c$ are the same parameters that describe 
the barrier shape (equation~\ref{gifbar}).  
The solid curve shows the predicted large scale Eulerian bias 
relation (with $a=0.707$, $b=0.5$ and $c=0.6$); it produces an 
upturn at the low mass end that is similar to the one seen in 
Jing's simulations.  
(In practice, the mass functions in the initial scale free simulations 
differ slightly from the GIF mass function.  So, strictly speaking, 
the bias relation should be computed using the values of $a$, $b$ and 
$c$ associated with the actual mass function in the scale free 
simulations.  Since this difference is small, we have not pursued 
this further.)

We end this section with a brief comparison of the ellipsoidal 
collapse bias relation with that in simulations which started from 
realistic initial power spectra.  Sheth \& Tormen (1999) showed that 
in the GIF simulations of SCDM, $\Lambda$CDM and OCDM models, the 
bias relation for haloes which are defined at $z_{\rm form}$ and are 
observed at $z_{\rm obs}=z_{\rm form}$ could be rescaled to produce 
a plot that was independent of $z_{\rm form}$ (see their Fig.~4).  
The symbols in Fig.~\ref{gifbias} show this rescaled bias relation 
for $z_{\rm form}=0$, 1, 2, and 4 (filled triangles, open squares, 
filled circles, and open circles, respectively).  
The dashed curves show the standard spherical collapse prediction, 
the dotted curves show the bias relation associated with the 
peak background split, and the solid curves show the ellipsoidal 
collapse prediction.  These GIF simulations span a smaller 
range in $\delta_{\rm sc}/\sigma$ than Jing's $n=-0.5$ scale free 
runs.  Over this smaller range, the peak background split formula 
and the moving barrier prediction are both in good agreement with 
the simulations.  

\section{Discussion}\label{conclude}
The mass function measured in simulations (equation~\ref{giffit}) 
is different from that (equation~\ref{oldps}) predicted by 
Press \& Schechter (1974) and by the excursion set approach of 
Bond et al. (1991) and Lacey \& Cole (1993).  If a model does not 
predict the mass function accurately, then the other model 
predictions, such as the large scale halo-to-mass bias relation, 
will also be inaccurate (e.g. Sheth \& Lemson 1999a; 
Sheth \& Tormen 1999).  It is important that a model describe 
both these statistical quantities accurately if the mass dependence 
of the abundance and spatial correlations of objects are to 
provide useful constraints on cosmological parameters (e.g. 
Mo, Mao \& White 1999; Arnouts et al. 1999; Moscardini et al. 1999).  
Since the excursion set approach allows 
one to make many analytic estimates about the evolution of 
hierarchical clustering relatively easily, it is worth 
modifying the original model so that it reproduces the simulation 
mass function.  The hope is that, if it predicts this accurately, 
the other predicted quantities will also be accurate.  

All predictions of the excursion set approach are based on solving 
problems associated with the time which passes before a particle 
undergoing Brownian motion is first absorbed onto a barrier.  
The predicted mass function depends on the height of the absorbing 
barrier as a function of random walk time.  Therefore, it is crucial 
to model this height accurately.  
Bond et al. (1991) argued that a barrier of constant height 
is associated with the dynamics of spherical collapse.  
Section~\ref{moving} of the present paper showed that combining the 
ellipsoidal collapse model for the dynamics with the assumption that 
the initial fluctuation field was Gaussian produces a barrier shape 
that is not constant (equation~\ref{elcolbar}).  Rather, it has a 
shape that is very similar to that which is necessary to 
produce a mass function like the one in numerical simulations 
(equation~\ref{gifbar}): it increases with decreasing mass.  

Our discussion of the excursion set approach in Section~\ref{model} 
allowed us to demonstrate that the inclusion of ellipsoidal dynamics 
(i.e., requiring that less massive objects be more overdense to 
collapse by a given time) reduces dramatically the scatter between 
the halo mass predicted by the theory and that which a halo 
actually has in simulations (Figs.~\ref{random}--\ref{errors}).  
That is, we showed explicitly that the ellipsoidal collapse, 
moving barrier, excursion set predictions work well on an 
object-by-object basis.  We then used the barrier crossing 
statistics of independent unconstrained random walks to provide 
an estimate of the halo mass function.  
Providing a more exact relation between the first crossing 
distribution of such walks and the mass function is the subject 
of ongoing work.  Even in this simple approximation, however, we 
argued that the approach also works well in a statistical sense.  
It predicts a mass function that has the same shape as the one in 
the simulations.  In addition, in contrast to the constant, spherical 
collapse barrier, our moving ellipsoidal collapse barrier predicts 
a large scale halo-to-mass bias relation (equation~\ref{lagbias}) 
that is similar to the one measured in simulations, even at the 
low mass end (Figs.~\ref{scfbias} and~\ref{gifbias}).  

We are not the first to consider the effects of non-spherical 
dynamics on the shape of the mass function of bound objects.  
Whereas Monaco (1995; 1997a,b), Audit, Teyssier \& Alimi (1997) 
and Lee \& Shandarin (1998) have studied models in which the 
initial deformation tensor is used to compute approximations 
to the collapse time, Bond \& Myers (1996) combined the 
information contained in the deformation tensor with the 
ellipsoidal collapse model to estimate the epoch of collapse.  
With the exception of Monaco, who assumed that virialization 
is associated with collapse of a single axis, all the other 
authors agree that it is the collapse of all three axes that is 
more relevant.  We agree.  As a result of his definition, Monaco 
found that the `moving' barrier should decrease, rather than 
increase, with decreasing mass.  One consequence of this is 
that if the barrier has the shape required by Monaco, then the 
inclusion of ellipsoidal dynamics would increase rather than 
decrease the scatter in our Fig.~\ref{centres}, relative to 
that associated with the `standard' spherical collapse model.  

We feel that our analysis incorporates some but not all of 
the various useful results derived by the authors cited above.  
For example, we could have computed the mass function following 
the `fuzzy' threshold approach of Audit, Teyssier \& Alimi (1997) 
and Lee \& Shandarin (1998).  In this approach, the `standard' 
spherical model corresponds to one in which all regions denser 
than a certain value $\delta_{\rm sc}$ collapse:  
$p({\rm collapse}|\delta)$ is a step function.  
Audit, Teyssier \& Alimi and Lee \& Shandarin provide various 
different definitions of this collapse probability, which are all 
motivated by combining approximations to non-spherical dynamics 
with the statistics of the initial shear field.  Figs.~2 and~3 of 
those papers show that, in such models, the probability of collapse 
is not a sharp step function.  For our definition of collapse, 
\begin{displaymath}
p({\rm collapse}|\delta) = \int_0^\infty {\rm d}e\!\! \int_{-e}^e 
{\rm d}p\ g(e,p|\delta)\ \Theta\bigl(\delta-\delta_{\rm ec}(e,p)\bigr)
\end{displaymath}
is not a step function either.  It is fairly straightforward to 
compute this probability using the results given in 
Section~\ref{moving}.  However, we feel that the excursion set 
approach allows one to estimate many more useful quantities than 
this fuzzy threshold approach.  This is why we have chosen to use 
our formula for $\delta_{\rm ec}(e,p)$ to compute a moving barrier 
shape, rather than to pursue the fuzzy threshold approach further.  

Another place where we could have done a more detailed 
calculation but did not is in relating mass and ellipticity.  
We used equation~(\ref{empv}) to provide a deterministic relation 
between $\sigma(m)$ and $e$, whereas there is considerable
scatter around this relation.  The authors cited above describe 
various methods for incorporating the effects of this scatter.  
In principle, we could apply any of their methods to our definition 
of collapse, and so include the effects of the scatter around the 
relation we use for translating $\delta_{\rm ec}(e,p,z)$ of 
Fig.~\ref{dcplot} to the moving barrier shape $B(\sigma,z)$ of 
equation~(\ref{elcolbar}).  [For example, equations~(24) and~(28) 
of Audit, Teyssier \& Alimi (1997) provide what is essentially 
their formula for what we call $B(\sigma,z)$, and their 
equation~(29) is an estimate for the scatter.]  
Whereas this might allow one to include the effects of the 
stochasticity resulting from a Gaussian fluctuation field more 
accurately (and so might allow one to reduce the scatter in 
Fig.~\ref{centres}), this increase in rigour is at the cost 
of making the other predictions associated with the excursion set 
model more difficult to compute.  
This is why we have not pursued this further.  

In this respect, our approach is more practical than rigorous.  
Because we are less careful than others about the exact stochasticity 
and dynamics, our approach (to provide an accurate fitting function 
to the barrier shape) is, perhaps, easier to implement.  
Indeed, we think it important to stress that, while it is 
reassuring that the barrier shape associated with the GIF mass
function can be understood within the context of a slightly more 
sophisticated treatment (than the spherical model) of the dynamics 
of collapse, the various other predictions of the excursion set model 
(the conditional mass function, the forest of merger history trees, and 
the nonlinearity and stochasticity of the halo-to-mass bias relation) 
are sufficiently useful, and sufficiently easy to make once the 
barrier shape is known, that they are worth making, using the fitting 
function of equation~(\ref{gifbar}), whether or not a more careful 
analysis of the dynamics of collapse and the stochasticity of the 
initial fluctuation field yields exactly the same barrier shape.  
The results presented in Section~\ref{apply} provide sufficient 
justification for using the barrier shape in this way.  
Making more such predictions is the subject of work in progress.  

Before concluding, we should mention that our moving barrier approach 
suggests that less massive objects at a given time must form from 
regions which are initially more overdense than the regions from 
which the more massive objects formed.  
This is in the same qualitative sense as the relation between mass 
and central-concentration that is measured for evolved halo density 
profiles (Navarro, Frenk \& White 1997).  These authors argue that 
less massive haloes are more centrally concentrated because, 
on average, the mass of less massive haloes was assembled earlier, 
at a time when the universal background density was higher.  
Our results suggest that at least some of this relation is built in.  

\section*{Acknowledgments}
Thanks to Tom Theuns for discussing how our formulae for 
ellipticity and prolateness are related to the Bardeen et al. 
formulae for peaks.

\appendix
\section{Gaussian random fields}\label{grfs}
Consider a Gaussian random field smoothed on scale $R_{\rm f}$.  
Let $\sigma(R_{\rm f})$ denote the rms fluctuation of the smoothed 
field.  Any position in this field has an associated perturbation 
potential, the second derivatives of which define what, in the 
Zeldovich approximation, is called the deformation tensor.  
Let $\lambda_1\ge \lambda_2\ge \lambda_3$ denote the 
eigenvalues of this tensor.  Different positions in the smoothed 
field will have different $\lambda_i$s.  The probability 
$p(\lambda_1,\lambda_2,\lambda_3)$ that the eigenvalues are 
$\lambda_1\ge \lambda_2\ge \lambda_3$, in that order, is 
\begin{eqnarray}
p(\lambda_1,\lambda_2,\lambda_3) &=& {15^3\over 8\pi\sqrt{5}\,\sigma^6}\ 
\exp\left(-{3 I_1^2\over \sigma^2} + {15 I_2\over 2\sigma^2}\right)
\times \nonumber \\
&&\qquad (\lambda_1-\lambda_2)(\lambda_2-\lambda_3)(\lambda_1-\lambda_3),
\label{lambdas}
\end{eqnarray}
where $\sigma\equiv \sigma(R_{\rm f})$, 
$I_1 \equiv \lambda_1 + \lambda_2 + \lambda_3$, and 
$I_2 \equiv \lambda_1\lambda_2 + \lambda_2\lambda_3 +\lambda_1\lambda_3$ 
(Doroshkevich 1970).  
In the linear regime, the initial density fluctuation is 
$\delta$, and because it is related to the potential by 
Poisson's equation, $\delta\equiv I_1$.  
By integrating $p(\lambda_1,\lambda_2,\delta-\lambda_1-\lambda_2)$ 
over $(\delta-\lambda_1)/2\le \lambda_2\le \lambda_1$, and then 
over $\delta/3\le \lambda_1\le\infty$, where the limits of 
integration follow from the fact that the eigenvalues are ordered, 
it is straightforward to verify that the distribution of 
$\delta$ is Gaussian, with variance $\sigma^2$.  
Also, since $\sigma^2\ll 1$ in linear theory, $|\delta|\ll 1$ 
almost surely, so the smoothing scale $R_{\rm f}$ has an associated 
mass $M\propto R_{\rm f}^3$.  

It is usual to characterize the shape of a region by its 
ellipticity, $e$, and prolateness, $p$, where  
\begin{equation}
e = {\lambda_1-\lambda_3\over 2\,\delta},\qquad{\rm and}\qquad 
p = {\lambda_1+\lambda_3-2\lambda_2\over 2\,\delta} 
\label{eandp}
\end{equation}
(e.g., Bardeen et al. 1986).  If we use the $\lambda$s from the 
formulae above, then the $e$ and $p$ values we obtain are those 
associated with the potential, rather than the density field.  
(So our $e$ and $p$ are what Bond \& Myers 1996 denoted $e_v$ 
and $p_v$, and they are not the same as what Bardeen et al. 1986 
call $e$ and $p$.)  The ordering of the eigenvalues means that 
$e\ge 0$ if $\delta>0$, and $-e\le p\le e$.  A spherical region 
has $e=0$ and $p=0$.  
Using equation~(\ref{eandp}) in Doroshkevich's formula allows 
one to write down the distribution of $e$ and $p$ given $\delta$.  
Let $g(e,p|\delta)\,{\rm d}e\,{\rm d}p$ denote this distribution.  
Then 
\begin{equation}
g(e,p|\delta) = {1125\over \sqrt{10\pi}}\,e\,(e^2-p^2)\,
\left({\delta\over\sigma}\right)^5
{\rm e}^{-{5\over 2}{\delta^2\over\sigma^2}(3 e^2 + p^2)},
\label{gepd}
\end{equation}
where we have used the fact that converting from 
${\rm d}\lambda_1{\rm d}\lambda_2{\rm d}\lambda_3$ to 
${\rm d}\delta\,{\rm d}e\,{\rm d}p$ introduces a factor of $2/3$ 
(Bardeen et al. 1986).
It is easy to verify that integrating this over $-e\le p\le e$, 
and then over $0\le e\le\infty$ gives unity, provided $\delta>0$.  
For all $e$, this distribution peaks at $p=0$.  
When $p=0$, the maximum occurs at 
\begin{equation}
e_{\rm mp}(p=0|\delta) = (\sigma/\delta)/\sqrt{5}.
\label{empv}
\end{equation}
This provides a monotonic relation between $e_{\rm mp}$ 
and $\delta/\sigma$.  Also, $e_{\rm mp}\to 0$ as 
$\delta/\sigma\to\infty$:
for a given $R_{\rm f}$ denser regions are more likely to be 
spherical than less dense regions, whereas, at fixed $\delta$, 
larger regions are more likely to be spherical than smaller ones.  
In general, the most probable shape of a randomly chosen region 
in a Gaussian random field is triaxial, with $p\approx 0$.  
Therefore, since we are interested in objects that form from 
Gaussian fluctuations, we must study the collapse of ellipsoids 
with $e\ge 0$.  Requiring that the initial regions of interest 
be peaks does not change these qualitative conclusions, 
though there are quantitative differences.  
Equations~(7.6) and~(7.7) of Bardeen et al. (1986) give the 
expressions corresponding to $g$ and $e_{\rm mp}$ for peaks 
(but note that their expressions are for the density, rather 
than the potential field).

\end{document}